\documentclass[pdflatex,sn-mathphys]{sn-jnl}
\jyear{2023}%

\usepackage{float}
\usepackage{etoolbox}
\usepackage{placeins}
\usepackage{amsmath}

\makeatletter
\patchcmd{\ps@headings}
{\hbox to \hsize{\hfill Springer Nature 2021 \LaTeX\ template\hfill}}
{\hbox to \hsize{}}
{}
{}
\patchcmd{\ps@headings}
{\hbox to \hsize{\hfill Springer Nature 2021 \LaTeX\ template\hfill}}
{\hbox to \hsize{}}
{}
{}
\patchcmd{\ps@titlepage}
{\hbox to \hsize{\hfill Springer Nature 2021 \LaTeX\ template\hfill}}
{\hbox to \hsize{}}
{}
{}
\makeatother

\begin{document}

\title[Goodrich et al., 2023]{Imaging of X-ray Pairs in a Spontaneous Parametric Down-Conversion Process}

\author[1]{Justin C. Goodrich}
\author[2]{Ryan Mahon}
\author[1]{Joseph Hanrahan}
\author[2]{Monika Dziubelski}
\author[2]{Raphael A. Abrahao}
\author[3]{Sanjit Karmakar}
\author[4]{Kazimierz J. Gofron}
\author[1]{Thomas Caswell}
\author[1]{Daniel Allan}
\author[1]{Lonny Berman}
\author[1]{Andrei Fluerasu}
\author[3]{Andrei Nomerotski}
\author[3]{Cinzia DaVià}
\author[1,5]{Sean McSweeney}

\affil[1]{National Synchrotron Light Source II, Brookhaven National Laboratory, Upton, NY 11973}
\affil[2]{Physics Department, Brookhaven National Laboratory, Upton, NY 11973}
\affil[3]{Department of Physics and Astronomy, Stony Brook University, Stony Brook, NY 11790}
\affil[4]{Spallation Neutron Source, Oak Ridge National Laboratory, Oak Ridge, TN 37831}
\affil[5]{Biology Department, Brookhaven National Laboratory, Upton, NY 11973}

\abstract{\textbf{} 
Spontaneous parametric down-conversion is a vital method for generating correlated photon pairs in the visible and near-infrared spectral regions; however, its extension to X-ray frequencies has faced substantial barriers. Here, we present an advancement in correlated X-ray pair generation and detection by employing a two-dimensional pixelated detector to obtain the first direct image of the pair distribution. Our study explores and directly visualizes the down-conversion process, revealing the characteristic ring structure of coincident photon pairs and demonstrating robust spatial correlations. A significant finding is the observation of energy anti-correlation, achieved at an unprecedented rate of approximately 4,100 pairs/hour, far exceeding previous reports in the literature. We believe these results represent a significant leap in X-ray quantum imaging, unlocking the potential for enhanced imaging of biological materials with reduced doses and broadening the applicability of X-ray quantum optical technologies.
}



\maketitle

The ability to generate, manipulate, and detect single photons has paved the way for many advances in quantum information science and technology. In particular, one of the most commonly used techniques to generate correlated single
photon pairs is spontaneous parametric down-conversion (SPDC)~\cite{SPDC_general,mosley2008heralded,grice2001eliminating,Brianna2023_SPDC}.
SPDC sources have revolutionized quantum physics~\cite{Kwiat,LoopholefreeBell_Vienna_TES,Loopholefree_Bell_NIST_nanowire,OAMentanglement} and quantum technologies~\cite{dowling&milburn,quantum_cnot,quantum_micros_HOM,moreau2019qimaging,Pittman,harder2013optimized}. In an SPDC process, a photon from a pump beam is absorbed in a material which results in the emission of a pair of single photons. SPDC is a nonlinear optical process with probabilistic characteristics that necessitates meticulous phase matching to satisfy the conservation of momentum. The down-converted state is dominated in number (Fock) basis by the $\lvert1,1\rangle$ state~\cite{SPDC_general}. The SPDC technique of correlated single photon generation is ubiquitous in the visible and infrared domains, however, extending this approach into X-ray wavelengths has remained a persistent challenge. Specifically, the low conversion efficiency in the most commonly used material, single crystal diamond, and the lack of fast, energy-resolving X-ray detectors have restricted advances in employing the X-ray SPDC process~\cite{Freund2,Eisenberger,Freund,Yoda,Adams,Adams2001,adams2013x_review,Sofer2019,Li2017,Schori2017,Schori2018-xv,Shwartz,Borodin,wong2021prospects}.

X-ray SPDC was initially demonstrated in the early 1970s using an X-ray tube source~\cite{Freund2,Eisenberger}, following theoretical predictions of the phenomenon in 1969~\cite{Freund}. In the early 2000s, X-ray SPDC was achieved with a diamond crystal as a nonlinear medium using a synchrotron source as the X-ray pump~\cite{Yoda,Adams,Adams2001}. However, these experiments produced limited coincidence count rates, on the order of a few coincidences per hour. Other experiments achieved coincidence count rates of approximately 90 photons per hour by working in a transmission Laue geometry instead of the reflected Bragg geometry~\cite{Shwartz}, and a more recent experiment reported coincidence rates of approximately 317 pairs per hour by using advanced background suppression methods~\cite{Borodin}.

In this work, we achieve an important milestone in X-ray SPDC, presenting record rates of down-converted pairs, an order of magnitude higher than the next best report in the literature~\cite{Borodin}. We also demonstrate precision measurements of the resulting SPDC spatial distribution and pair energy anti-correlations. Our advancements set the stage for a transformative objective: the imaging of radiation-sensitive biomaterials~\cite{BioDamage,BioDamage2} with reduced radiation dosage by utilizing entangled X-rays. This approach provides a glimpse into the future of X-ray quantum and ghost imaging by leveraging quantum principles to enhance precision and signal-to-noise ratio beyond the capabilities of classical methods~\cite{wong2021prospects,QuantumIllum,QuantumIllum2,QuantumIllum3,moreau2019qimaging}. By merging fundamental physics with practical applications, our work potentially opens a new avenue for investigating foundational questions in quantum physics, thereby bridging the gap between theoretical concepts and real-world applications. Even if our present focus is on the X-ray SPDC process itself, this report represents a critical step toward realizing these transformative imaging techniques.

\subsection*{SPDC Photon Pair Generation and Detection}

\begin{figure}[htp!]
\centering
\includegraphics[width=1\textwidth]{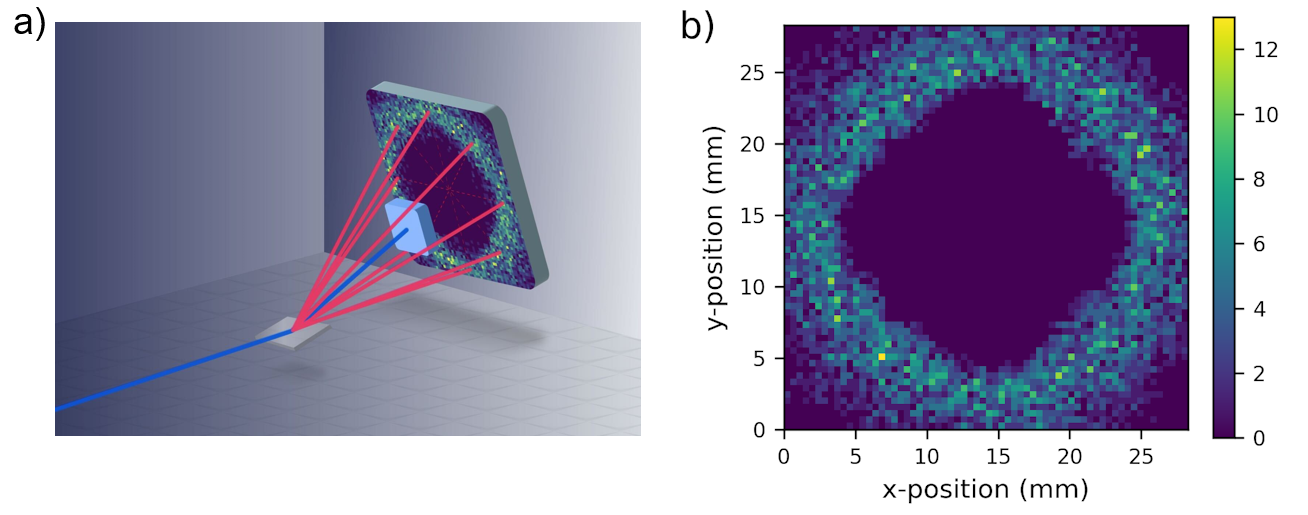}
\caption{\textbf{Conceptual experimental schematic and the final observed SPDC structure.} \textbf{a,} A conceptual schematic of the experiment, which shows that the Bragg reflection of a pump incident upon (111) diamond, with a small detuning, can generate down-converted X-ray pairs around the diffracted pump. A tantalum beamstop is played to obscure flooding the detector with diffracted pump photons. \textbf{b,} The detected SPDC photons after isolation from background scattering using a robust filtering process. The results are from a one hour exposure at $\Delta\theta$ = 0.022$^\circ$ with a total count of 4,145 photon pairs. Analyses of the spatial and spectral structures follow.}
\label{fig1}
\end{figure}

A conceptual schematic (Fig. \ref{fig1}a) of our experiment highlights the key details: coherent hard X-rays from a synchrotron light source perform nonlinear diffraction with a diamond crystal, which results in the emergence of a ring structure of correlated photon pairs. Unlike optical SPDC, which typically relies on birefringence, X-ray SPDC materials often lack strong nonlinearities at the relevant wavelengths; in this work, we rely on nonlinear Bragg diffraction using diamond's (111) reflection and the recombination of virtual intermediate states for generation of the down-converted photons~\cite{Freund,Eisenberger}.

When employing Bragg diffraction using a diamond single crystal, the reflection's Bragg angle is detuned slightly to set the phase matching condition between pump photons, the reciprocal lattice vector, and the two down-converted single photons (often called the ``signal'' and ``idler'' photons in the quantum optics community). This guarantees momentum conservation between the down converted photons and the pump~\cite{Freund,Eisenberger}, resulting in the formation of a diffuse ring structure of photons centered about the diffracted pump reflection. This signal is obscured by a large background of scattered 15 keV pump photons incident upon the detector, from which it must be isolated. The final structure of the X-ray SPDC phenomenon (Fig. \ref{fig1}b) is visible after a robust data filtering and a photon pairing procedure, discussed in detail below, are employed to remove as much background as possible. This constitutes both the first imaging of the X-ray SPDC ring with an area detector and the highest count rates to date, approaching 4,000 pairs per hour.

We use three independent observables, enabled by the new Lynx T3 detector~\cite{timepix3}, to distinguish the SPDC pair signal from background noise: the time-over-threshold (ToT) of events, which is correlated to the photon energy; the nanosecond-scale time-of-arrival (ToA) of events; and their positional (x, y) pixel coordinate information. Refined selections based on conservation of energy and conservation of momentum, determined from the positional information, can be performed to dramatically increase the SPDC signal-to-noise ratio (SNR).

\begin{figure}[htp!]
\centering
\includegraphics[width=1\textwidth]{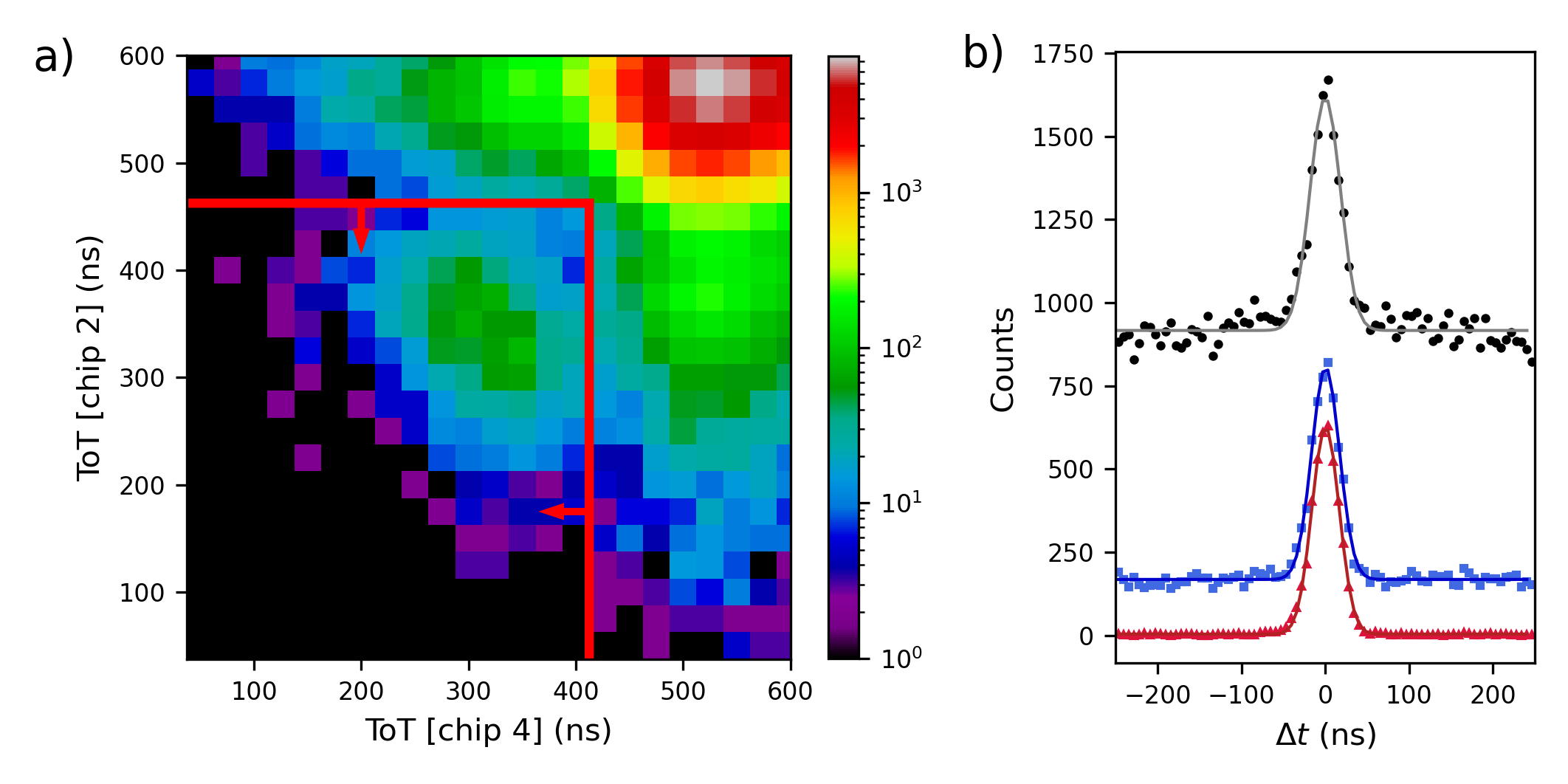}
\caption{\textbf{Experimental observables leading to the SPDC pair selections.} \textbf{a,} The ToT of photons in the top left chip against the ToT of its paired photon in the bottom right chip, by minimal time difference. In this plot, the ToT selections were raised to a value of 600 ns, with values of the selection thresholds used for the SPDC search indicated by red arrows. The logarithmic scale highlights that the data are dominated by high ToT pump photons in the range of 500-600 ns, with a small SPDC signal at lower ToTs. \textbf{b,} Histograms of the time coincidence measurements with Gaussian fits, at various points in the data filtering process, using the ToT selection thresholds indicated with the red arrows in Fig. 2a. The histograms have a bin size of 6.25 ns (4 ToA bins). Black circles with gray line (top): measured pairs by minimal time difference after only ToT filtering with a Gaussian distribution fit ($N = 5,149 \pm 249$ pairs with SNR = 0.76; $\sigma_{\Delta t} = 18.1$ ns). Blue squares with dark blue line (middle): ToT-filtered data after selecting pairs that show conservation of the pump energy with a Gaussian distribution fit ($N = 4,281 \pm 97$ pairs with SNR = 3.80; $\sigma_{\Delta t} = 16.5$ ns). Red triangles with dark red line (bottom): ToT-filtered data after selecting pairs that show both conservation of the pump energy and momentum with a Gaussian distribution fit ($N = 4,126 \pm 39$ pairs with SNR = 182.7; $\sigma_{\Delta t} = 16.3$ ns).}
\label{fig2}
\end{figure}

ToT selections (Fig. \ref{fig2}a) are employed where events above certain ToTs are dropped, removing a significant number ($\sim$98\%) of recorded events from the raw data. The ToT selection thresholds were optimized for each of the four Timepix3 chips in the detector assembly, targeting the valley of the ToT spectra between 7.5 keV and 15 keV. This optimization preserves as much of the SPDC signal as possible while reducing the background at 15 keV.  

After ToT filtering, photons are paired together based on their time correlation (by minimum time difference). Two photons generated in an SPDC process should be coincident in time when reaching the detector, producing a peak with a small time difference window centered at $\Delta t$ = 0 ns (Fig. \ref{fig2}b). Three datasets are shown with successive selections: pairs found after only the ToT selections (black circle data/gray fit), pairs found with the additional constraint that the photons conserve the pump energy $E_i + E_s = E_{pump}$ (blue square data/dark blue fit), and pairs found with the third constraint that photons conserve the pump momentum $\Vec{\rho}_i + \Vec{\rho}_s = \Vec{\rho}_{pump}$ (red triangle data/dark red fit). The noticeable background in the $\Delta$t distributions prior to the final energy and momentum constraints is due to random coincidences of the pump X-rays, corresponding to a total photon rate of $\sim$7.3 kHz with our ToT selections, and a possible but rare $\lvert2,2\rangle$ down-converted state. Using the momentum and energy constraints reduces this background to a very low level and enables a high SNR greater than 100 (20 dB). These latter constraints are described in detail in the following section. We fit the peaks with scaled and shifted Gaussian distributions, which give $\sigma_{\Delta t}$ $\approx$ 16.3 ns, corresponding to the time resolution per X-ray photon pair. Assuming an equal contribution of each photon in the pair to the time uncertainty, we obtain 16.3 ns/$\sqrt{2} \approx$ 11.5 ns for the corresponding temporal resolution per photon. We discuss the time resolution further in the Methods section. 

\subsection*{Spatial Properties and Energy Anti-Correlation}

The theoretical treatments of this phenomenon from Freund and Levine~\cite{Freund} and Eisenberger and McCall~\cite{Eisenberger} indicate that the SPDC of X-rays should be a non-degenerate process with a wide energy spectrum, of the form:

\begin{equation}
\eta_{\text{SPDC}}(E) \propto E(E_{pump} - E)
\end{equation}

where $\eta_{\text{SPDC}}$ is the probability of generating a photon of energy $E$ from a pump photon energy of $E_{pump}$ in the pair production process (see Supplemental Information for derivation from equation (8) in ~\cite{Eisenberger}). The generated photons may possess any energies that satisfy energy conservation:

\begin{equation}
E_i + E_s = E_{pump},
\end{equation}

and momentum conservation:

\begin{equation}
\Vec{\rho}_i + \Vec{\rho}_s = \Vec{\rho}_{pump}
\end{equation}

where the suffixes \textit{i} and \textit{s} refer to idler and signal photons, respectively. The conservation of momentum requires that the two photons in the pair produced in the SPDC process will be colinear with and on opposite sides of the Bragg peak, with distances from the peak position determined by their energies. The relationship between a single photon's energy \textit{E} and its emission angle $\alpha$ from the diffracted pump beam at 2$\theta$ is given by the following equation (see Supplemental Information for derivation from equation (7) in ~\cite{Eisenberger}):

\begin{equation}
E = \frac{E_{pump}}{\frac{\alpha^2}{2 \Delta\theta \sin 2\theta} + 1}
\end{equation}

where 2$\theta$ is the angle of the diffracted pump from the (111) diamond plane and $\Delta\theta$ is the detuning angle from the Bragg scattering $\theta$ angle. The factor $2\Delta\theta\sin2\theta$ is determined by the phase matching condition. The relationship between $\alpha$, the photon distance \textit{r} from the 2$\theta$ position on the detector plane, and the distance \textit{L} between the diamond and the detector is given by $tan(\alpha) = \frac{r}{L}$ (Extended Data Fig. \ref{efig1}), which can be used with equation (4) to determine the relationship between \textit{r} and \textit{E}. 

\begin{figure}[htp!]
\centering
\includegraphics[width=0.49\textwidth]{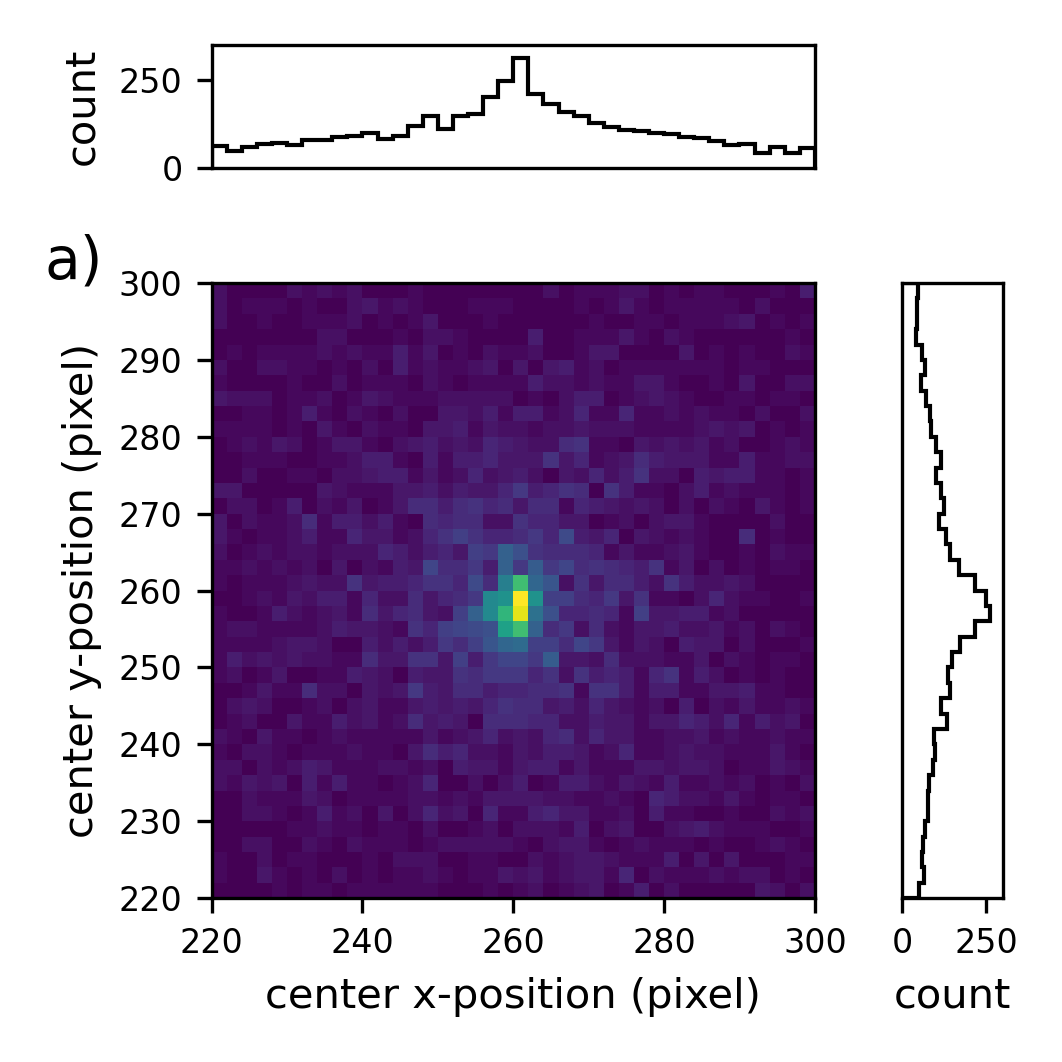}
\includegraphics[width=0.49\textwidth]{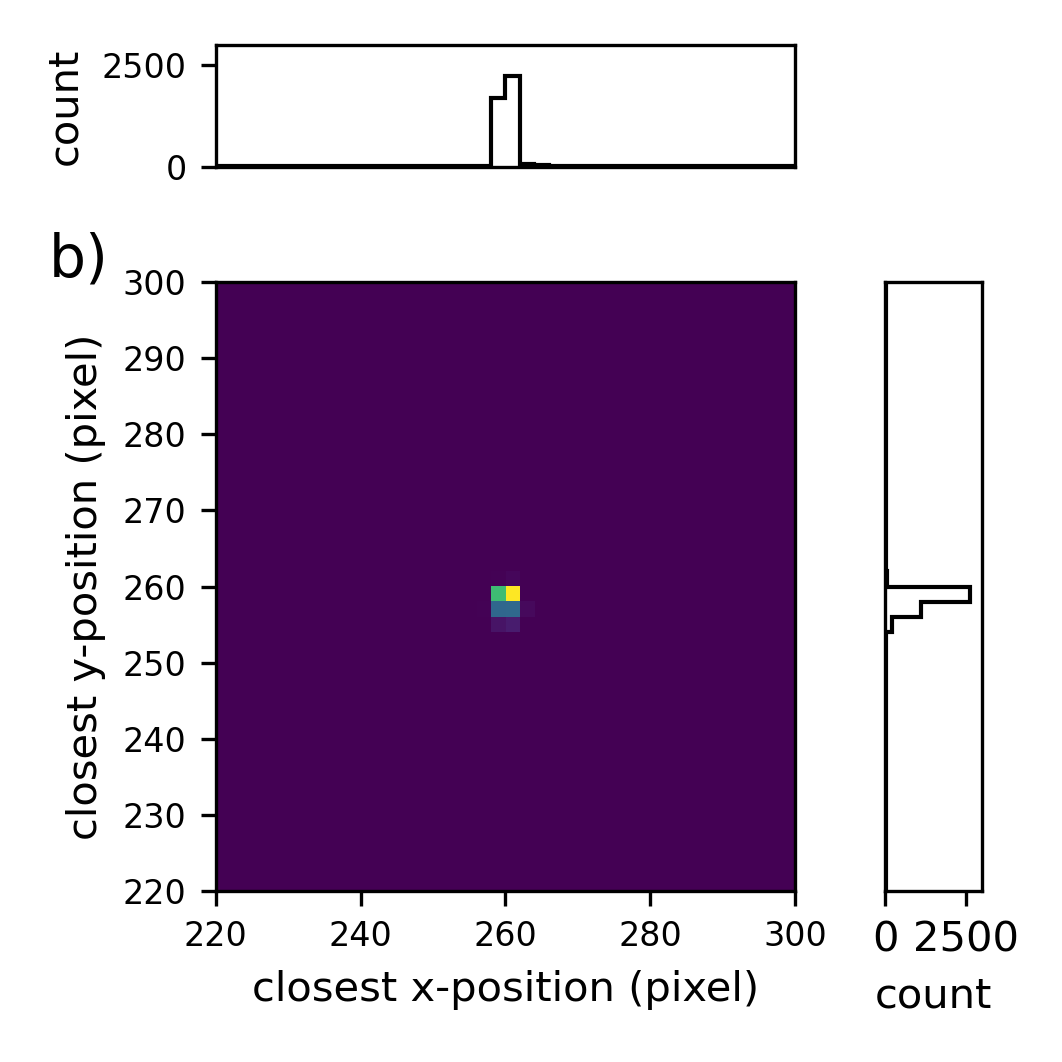}
\caption{\textbf{Analysis of the spatial location of pair center points and pair collinearity with the Bragg reflection.} \textbf{a,} Distribution of the average of each selected pairs' (x, y) pixel coordinates. The peak around pixels (260, 259) corresponds to the same location of the (111) Bragg reflection on the detector, before being masked by a tantalum beamstop. The histogram bin sizes are 2x2 pixels for better visibility. 1D histograms along each axis are also presented. \textbf{b,} Location on the line segment connecting paired photon locations that are closest to the Bragg peak location. Pairs within a radius of 4 pixels of the Bragg location at (260, 259) are considered to satisfy the conservation of momentum. The histogram bin sizes are 2x2 pixels for better visibility. 1D histograms along each axis are also presented.}
\label{fig3}
\end{figure}

The calculation of the center point (Fig. \ref{fig3}a) between pairs results in a peak around the detector pixels (260, 259). This corresponds to the measured location of the diffracted pump, consistent with the most likely scenario where $E_i = E_s = \frac{1}{2}E_{pump}$. Non-degenerate photon pairs cause a deviation from the degenerate center point. In those cases, the calculated center point is shifted towards the location of a lower energy X-ray, which has a larger emission angle $\alpha$ than its higher energy counterpart. The strong agreement between the calculated positions and the measured diffracted Bragg peak provides strong evidence that the selected pairs emerge from the SPDC process.

Calculating the closest position on the line segment connecting photons in a found pair with the Bragg location at (260, 259) provides a measure of the collinearity of the pairs with the pump (Fig. \ref{fig3}b). This measurement provides a background reduction mechanism, as the line segment for true SPDC pairs will pass through or close to the pump location due to conservation of momentum, whereas pairs formed from the background events are unlikely to be colinear with the pump. Pairs with a line segment that passes within a small distance of 4 pixels (see Extended Data Fig. \ref{efig2} for the impact and selection of this value) from the pump location are considered to satisfy the conservation of momentum. This criterion is used with conservation of energy, discussed next, to filter the data to generate the final ring (Fig. \ref{fig1}) and the red data points in the time coincidence plot (Fig. \ref{fig2}b), providing a substantial improvement in the SNR.

\begin{figure}[htp!]
\centering
\includegraphics[width=1\textwidth]{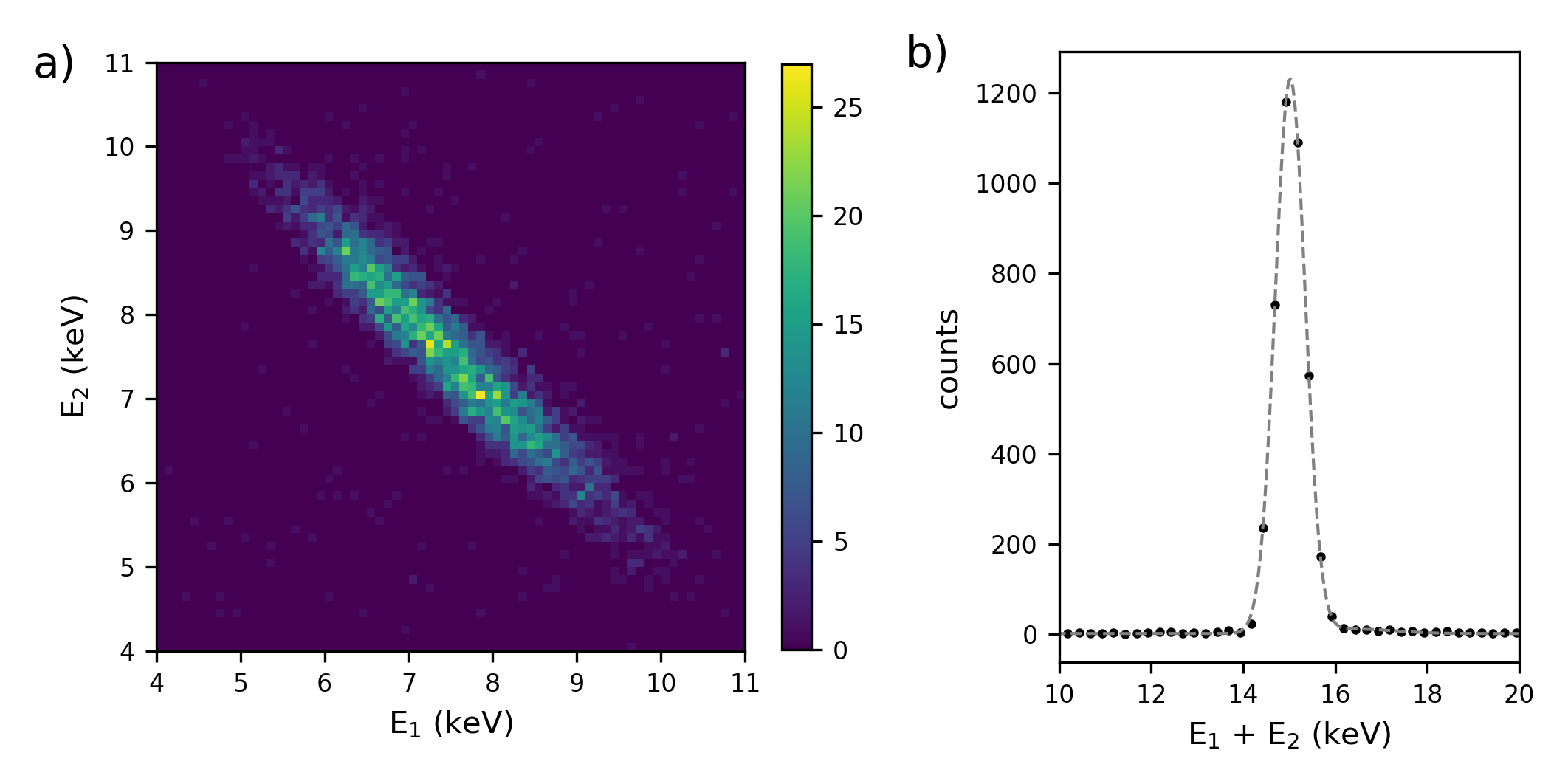}
\caption{\textbf{Anti-correlation of SPDC pair energies.} \textbf{a,} Distribution of SPDC X-ray radii, converted to energy using equation (3). A robust, linear energy anti-correlation is observed, indicating energy conservation. \textbf{b,} Distribution of the sum of the energies for both photons in each pair. The histogram bin size is 0.25 keV. Black circles indicate calculated values from the experimental data and the gray line is a Gaussian fit ($N = 4,145 \pm 39$ pairs with SNR = 340.0, $E_0 = 15.0$ keV, $\sigma_E$ = 0.33 keV).} 
\label{fig4}
\end{figure}

We can demonstrate the energy anti-correlation of the photons in the measured pairs by converting each photon's distance \textit{r} from the Bragg peak location into energy using equation (4). This calculation can be used to understand the spectral properties of our experimental setup. Comparing $E_i$ vs. $E_s$ (Fig. \ref{fig4}a) for all of the pairs selected with time coincidences within $\pm 2\sigma_{\Delta t}$ and using conservation of momentum selections shows that the energies of the two photons in a pair have anti-correlated energies, a fundamental property of the SPDC process~\cite{Kwiat, nomerotski2023timepix3}.

\FloatBarrier

The sum of the energies of the photon pairs, $E_i + E_s$ (Fig. \ref{fig4}b), shows a peak at 15.0 keV, the pump energy, consistent with the requirement of conservation of energy. A small background of paired events, constituted by scattered pump photons that make it through the selections, is present. Note that equation (4) only applies to SPDC-generated photons and does not apply to background pump photons which make it past selections and are inadvertently paired. Pairs with an energy sum of 15.0 keV $\pm$ $2\sigma_{E}$ are considered to conserve energy, which is used in conjunction with the conservation of momentum requirement as a selection parameter for the final ring structure (Fig. \ref{fig1}) and time coincidence histograms (Fig. \ref{fig2}).

Applying the conservation of energy and conservation of momentum restrictions to the data reveals a final total signal (SPDC) pair count of 4,145 $\pm$ 39 pairs/hour or $\sim$1.15 Hz; this value is what we take to be our measured SPDC rate. This rate is over an order of magnitude larger than the next highest rate reported in the literature (317 pairs/hour)~\cite{Borodin} and reveals the diffuse ring-like spatial structure generated by the X-ray SPDC process (Fig. \ref{fig1}b).

\subsection*{Detuning Angle Dependence}

\begin{figure}[htp!]
\centering
\includegraphics[width=0.6\textwidth]{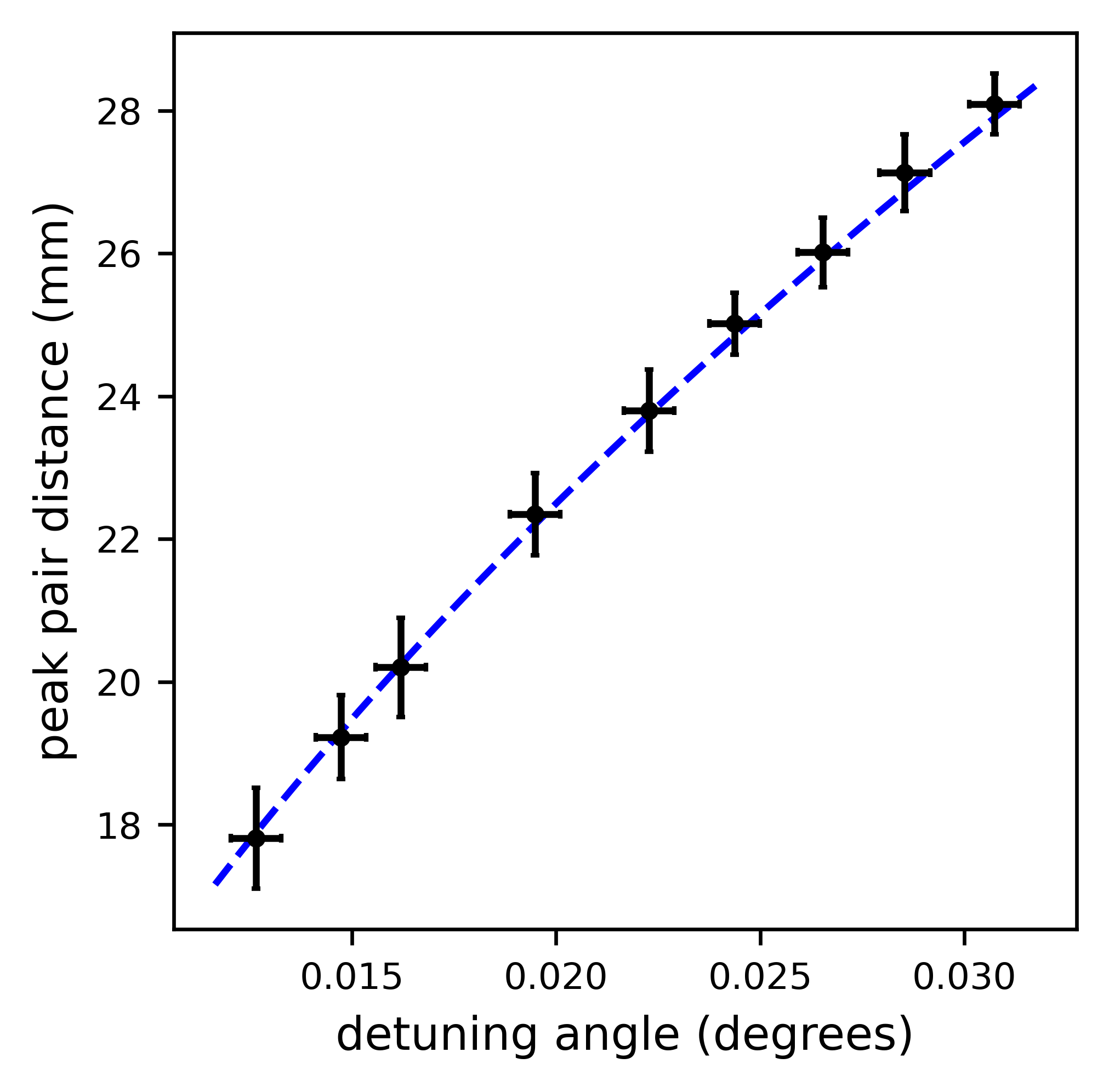}
\caption{\textbf{Dependence of the detuning angle on the peak distance between photon pairs in the SPDC structure.} The peak diameter is associated with the degenerate case where $E_i = E_s$. The experimental results (black circles, horizontal error bars associated with uncertainty of alignment to the peak position of the diamond (111) rocking curve with small vertical bars representing the standard deviation of the distance measurements) against theoretical predictions of the degenerate energy distance from equation (4) (dashed blue line).}
\label{fig5}
\end{figure}

The emission angles of the down-converted photons are determined by the detuning angle $\Delta\theta$. The results presented thus far are for a detuning angle of $\Delta\theta \approx 0.022^\circ$. To further validate the phase matching condition, we systematically varied this angle and accumulated corresponding datasets. The dependence of the measured degenerate photon pair distance on $\Delta\theta$ matches very well with theoretical predictions~\cite{Freund,Eisenberger} from the satisfaction of the phase matching condition (Fig. \ref{fig5}).

\FloatBarrier

This confirmation reinforces the fact that diamond is a robust nonlinear medium for the generation of SPDC correlated X-ray pairs. The detuning angle can be set or swept according to the desired application, allowing for precise control over the spatial/energy relationship of the down-converted X-rays, a facet very useful for quantum imaging applications and explored in our future publications. 

\subsection*{Detector Acceptance and Beamstop Obstruction}

A comprehensive model that aligns well with our observed experimental data (Fig. \ref{fig6}) can be developed by integrating the impact of the detector acceptance, beamstop obstruction, and ToT filtering with the theoretical energy distribution of SPDC single photons, equation (1). Equation (4) establishes a relationship between the energy (E) of an SPDC single photon and the angle $\alpha$ at which it emerges from the crystal, relative to the Bragg $2\theta$ direction; low-energy X-rays are emitted at larger angles, whereas high-energy X-rays nearing the pump energy have smaller emission angles.

Our experimental setup incorporated a 12.7 mm x 12.7 mm square beamstop, positioned close to the detector's center, and the detector's active region measures 28.3 mm x 28.3 mm. The distance from the diamond to the detector was 69.7 cm. This geometry inherently restricts the energy acceptance of our measurements. The effect of this phenomenon can be quantified by calculating the proportions of the circumferences of concentric circles, centered at the Bragg peak location, that are obscured by the beam stop or fall outside the detection plane. This geometry establishes the acceptance of our apparatus as a function of photon energy. High-energy SPDC photons exceeding $\sim$12 keV are partially obstructed by the beamstop, with complete suppression occurring at $\sim$14 keV. Conversely, X-rays with energies below $\sim$6 keV begin to not be accepted by the detector plane and fully disappear at $\sim$4 keV. If one of the two down-converted photons is not detected, the pair will not be found. This restriction shapes the observed spatial structure, causing the circular ring structure to appear more square-like in appearance. 

The other influential variable that shapes our observed photon energy distribution pertains to our ToT selection procedure. The ToT spectra of monochromatic photons generally follow a normal distribution, displaying a bandwidth on the order of approximately 125 ns, increasing with energy. The average value of the ToTs recorded from such an experiment can be used to coarsely map energy to a mean ToT (Extended Data Fig. \ref{efig3}). Although rudimentary, the ToT offers insight into estimating the likelihood that a recorded photon event possesses a specific energy. Events surpassing certain ToT selection thresholds are excluded to mitigate the considerable background of scattered 15 keV pump photons. Lower ToT selections yield a higher SNR but fewer pairs, whereas higher ToT selections lead to more pairs, albeit with significantly amplified background counts. An inadvertent result of this procedure is the elimination of certain SPDC photons, particularly high-energy ones, due to their exceeding the ToT selection threshold. This effect can be modeled with the following equation:

\begin{equation}
\eta_{\text{ToT}}(E,\text{ToT}_{\text{cutoff}}) = \frac{\int_{0}^{\text{ToT}_{\text{cutoff}}} \text{ToT}_{\text{count}}(E, \text{ToT}) \, d\text{ToT}}{\int_{0}^{\infty} \text{ToT}_{\text{count}}(E, \text{ToT}) \, d\text{ToT}}
\end{equation}

where $\eta_{\text{ToT}}$ represents the efficiency of the ToT filtering, $E$ is the photon energy, and $\text{ToT}_{\text{cutoff}}$ is the chosen threshold.

Both of these models provide detection efficiencies as a function of energy. Combining these curves with the conversion cross-section of SPDC X-rays in equation (1), and requiring both photons to be detected to form a pair, results in a thorough modeling of different facets of the experiment that impact the measured energy distribution of the SPDC process.

\begin{figure}[htp!]
\centering
\includegraphics[width=0.7\textwidth]{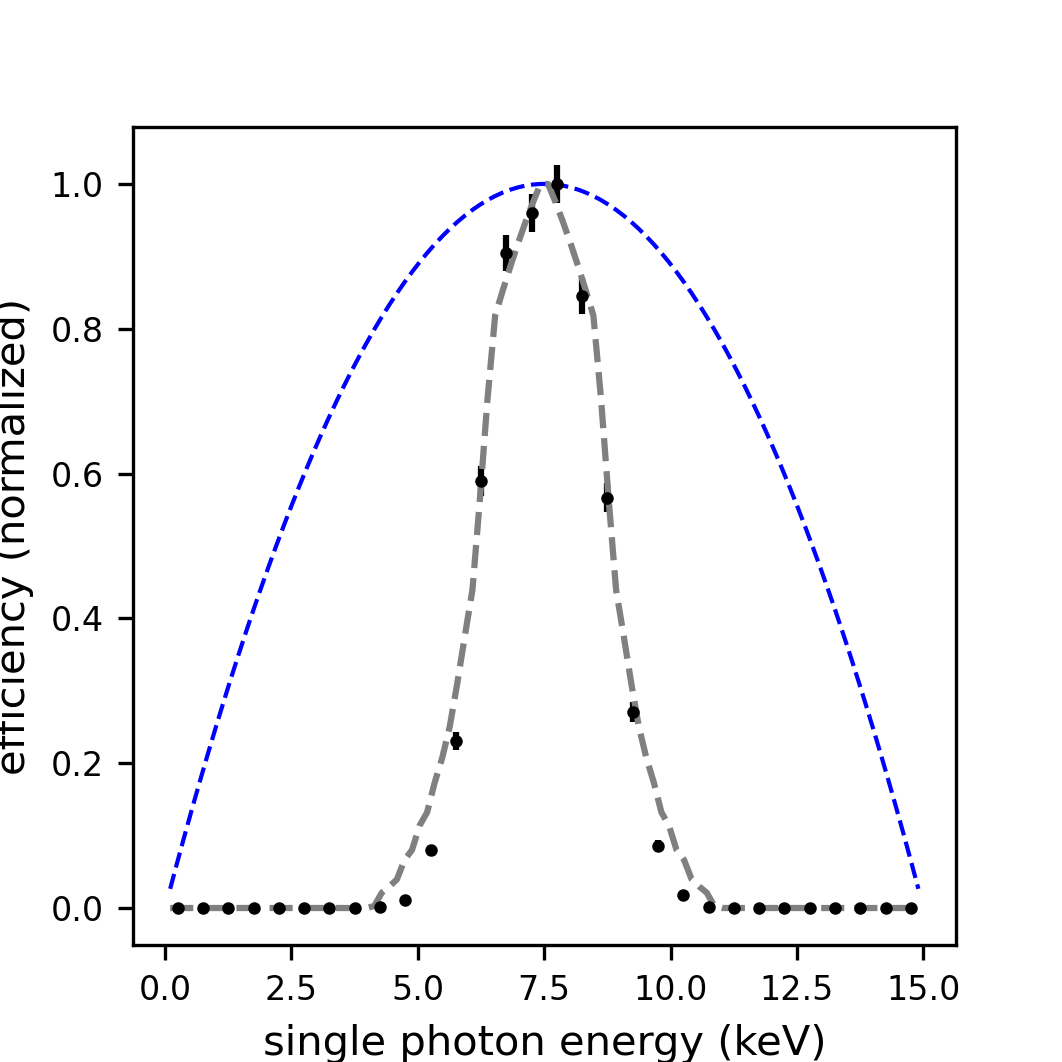}
\caption{\textbf{Experimental limitations of the SPDC energy measurements.} The theoretical predictions of conversion efficiency as a function of energy, which has the form of an inverted parabola with zeros at 0 energy and the pump energy, 15 keV (blue curve), along with the model of the geometrical effects from the detector acceptance and beamstop obstructions, combined with calculations of the effect of the ToT selections, multiplied by the theoretical predictions (gray curve). A histogram of SPDC single photon energies with error bars from the estimate of the standard deviation of the experimental counts (black circles) is presented to confirm the validity of the model. The histogram bin size is 0.5 keV.}
\label{fig6}
\end{figure}

We contend that the robust agreement between the experimental data and this comprehensive model (Fig. \ref{fig6}) confirms the validity of the theory, even with the constraints imposed by our apparatus's geometry in detecting the full spectrum of SPDC photon energies. This correlation serves as substantial evidence for the observation of non-degenerate X-ray pairs in an SPDC process. Additionally, the discernible energy anti-correlation within the spectral bandwidth permitted by our setup further supports this conclusion. A comparison of the spectral range measurable by our experimental setup with the theoretical efficiency of the process indicates that we are capable of detecting $\sim$30\% of the down-converted photon pairs.

\subsection*{Outlook}

The work presented here marks significant milestones in X-ray quantum optics, reporting the highest detection coincidence count rates to date, the first imaging of the SPDC ring, and a detailed characterization of properties for the down-converted X-rays, including their spatial and energy correlations. These advancements in count rates and the improved detection system have enabled us to investigate the spatial correlations of paired X-rays in an unprecedented way.

Further potential improvements are possible with the use of more advanced detectors with capabilities beyond Timepix3-based systems. In our case of hard X-rays, simultaneous direct measurements of time, energy, and position are achievable. Although the energy resolution of the Timepix3 detector through the ToT measurement is not optimal, a clear correlation of the measured ToT and the SPDC photon position is evident (Extended Data Fig. \ref{efig3}). This finding independently confirms the energy measurement through position information. A detector with improved energy resolution could refine these measurements and enable enhanced probing of multi-dimensional correlations of time, energy, and position in the X-ray SPDC process. Such detectors could significantly improve signal/background separation, imaging capabilities, and reduce the impact of ToT filtering on the detection of non-degenerate X-ray pairs.

Exploring other anticipated lines of investigation, such as the use of different diamond reflections or alternative materials for the down-conversion process, holds promise. The potential applications of X-ray quantum optics extend to biological and medical imaging techniques, aiming for superior image quality with reduced radiation doses. Our work could also offer an alternative to existing quantum microscopes~\cite{quantum_microscope,quantum_micros_HOM} and enable high-quality quantum ghost imaging in the X-ray regime. Ultimately, we believe that our research paves the way for numerous new lines of research and applications in X-ray quantum optics, enriching the field with innovative approaches and techniques.

\section*{Methods}

Our investigations were conducted at the 11-ID Coherent Hard X-ray Scattering (CHX) beamline of the National Synchrotron Light Source II (NSLS-II) facility at Brookhaven National Laboratory (BNL). The beamline was configured to produce a monochromatic X-ray beam set at 15 keV. The average input flux of the pump beam generated at the BNL’s NSLS-II beamline was approximately $10^{11}$ photons/second, with a polarization of 99\% in the horizontal direction and $\frac{\Delta E}{E} \approx 10^{-4}$. The dimensions of the 15 keV input beam were set to 50 $\mu$m (horizontal) x 50 $\mu$m (vertical). The incident X-ray beam was directed onto a (111) diamond single crystal with dimensions of 3 mm x 3 mm x 0.33 mm, grown by chemical vapor deposition from Element Six~\cite{element6}, which was used as the nonlinear medium for the generation of correlated photon pairs. A vacuum-pumped flight path beam pipe was installed between the diamond crystal and the detector surface. The total path distance from the crystal to the detector surface was 69.7 cm.

After standard height alignment procedures, the diamond crystal was oriented to the (111) Bragg reflection at a Bragg ($\theta$) angle of $11.576^\circ$. Multiple positions along the surface of the diamond were probed to find the area with the highest quality, using rocking curves ($\theta$ scans) to minimize dual peaks and full-width-at-half-max (FWHM). These measurements were performed using CHX's Eiger X 500K detector~\cite{Eiger}. The final position was measured to have a mosaic spread of $\sim$11 arcseconds. The resulting diffraction pattern peaked around a pixel coordinate of (260, 259) on a Lynx T3 detector, which was subsequently employed for measurement of the SPDC signal.

A tantalum beamstop was strategically positioned to obscure the direct Bragg reflection and nearby scattering. The Bragg alignment was then fine-tuned by a deviation of +$0.022^\circ$ to a $\theta$ angle of $11.598^\circ$ to meet the phase matching condition, a requirement for the SPDC process. This adjustment instantiated the production of correlated X-ray pairs at a peak output angle of $0.995^\circ$, an angle chosen to maximize the SPDC signal between the edge of the detector and the tantalum beamstop. The experimental schematic (Extended Data Fig. \ref{efig1}) illustrates the detailed configuration of our setup. 

To isolate the correlated X-ray photon pairs, we employed a Lynx T3 detector by Amsterdam Scientific Instruments, equipped with four Timepix3 readout application-specific integrated circuits (ASICs)~\cite{timepix3}. This detector has a 512 x 512 pixel array with a pixel size of 55 x 55 $\mu$m. This detector was chosen for its accuracy in measuring the time-of-arrival and energy deposition. Two photons generated in an SPDC process should be coincident in time when reaching the detector. However, nonlinearities in the timestamping process result in a ``timewalk"\cite{Timewalk}, whereby events that arrive at the same time might measure a time difference of up to 50 ns. This effect is observed in the $\sigma_{\Delta t}$ of the time difference histogram fits (Fig. \ref{fig2}). The detector was integrated into the NSLS-II CHX beamline's control system through the development of an appropriate EPICS areaDetector driver~\cite{AD,ADTimepix} and Python objects for integration with the Bluesky RunEngine environment~\cite{Bluesky}.

A minor fraction of hot pixels ($\approx$0.1\%) were masked to avoid recording substantial amounts of noise hits. This system allows for multihit functionality for each pixel, independent of others, along with a rapid readout bandwidth of 80 MPix/sec. Occasionally, photons excite charge in multiple adjacent pixels; to correct for this, a k-d tree algorithm was employed to cluster and ultimately centroid such events together. The 512x512 pixel array has two columns and two rows along the center (x and y = 255 and 256) of large, double-length (110 x 55 $\mu$m) pixels. This is corrected by inserting rows and columns of two dummy pixels between chips to preserve the correct distance between pixels.   

We calibrated the ToT dependence on X-ray energy by allowing the detection of beams of scattered X-rays, varying their energy from 6 to 15 keV in 0.5 keV steps (Extended Data Fig. \ref{efig3}a). We also cross-calibrated the ToT energy estimator using the spatial information of selected SPDC pairs (Extended Data Fig. \ref{efig3}b), which gave good agreement from the scattering calibrations. These two calibration techniques provide reliable and redundant ways to determine the X-ray energy, with the second approach only applicable for down-converted X-rays, but providing considerably more precise energy determination. The dual-calibration method ensures robust accuracy, leveraging spatial information as a complementary validation to the ToT measurements.

All data processing and analysis were conducted using custom Python code, developed in-house and computed on CHX's local resources. This tailored approach facilitated the meticulous processing of the raw Lynx T3 detector output, including corrections, thresholding, filtering, time coincidence searches, photon pair determinations, and analysis of spatial and energy correlations.

\subsection*{Impact of Selection Parameters}

The raw data recorded by the detector, prior to any selections, resemble a generic X-ray scattering pattern with no obvious SPDC signature (Extended Data Fig. \ref{efig4}a). The process to achieve the final SPDC structure (Fig. \ref{fig1}b) includes four critical steps: (1) ToT filtering to eliminate high-energy background, (2) selecting events within a small time window $\pm 2 \sigma_{\Delta t}$(Extended Data Fig. \ref{efig4}b), (3) selecting pairs with an energy sum corresponding to the pump energy of 15 keV $\pm 2 \sigma_{E}$ (Extended Data Fig. \ref{efig4}c), and (4) selecting pairs colinear with (within 4 pixels of) the pump location (Fig. \ref{fig1}b). An iterative approach of applying these selections can provide the best selection parameters to retain a high count rate while reducing the background.

The collinearity selection tolerance is the largest handle in  background suppression (Extended Data Fig. \ref{efig2}). A value of 4 pixels was chosen because it optimizes the SNR without a significant reduction in count; smaller values begin to cut SPDC photons that are slightly off center from the peak pump location due to the mosaic spread of the diamond, whereas higher values do not result in a significant change in the number of pairs but degrade the SNR due to background pairs which satisfy the looser colinear check.

Another essential aspect of our selection process is the possibility of precisely setting of the ToT selection thresholds using the method described. It should be noted that higher SPDC photon counts could be inferred with less conservative ToT filtering. We adjusted these thresholds in increments of 25 ns and observed that raising the thresholds led to a greater number of pairs identified through Gaussian distribution fits, albeit at the expense of lower SNRs. The thresholds utilized in the results presented were carefully chosen by calibrating the ToT distributions per chip, using 7.5 keV and 15 keV monochromatic photon scattering. The selected values represent an optimal balance between the high signal rate and SNR, as they allow for enhanced pair detection without a significant loss in signal quality.

\section*{Extended Data}

\begin{figure}[htp!]
\centering
\renewcommand{\figurename}{Extended Data Fig.}
\setcounter{figure}{0}
\includegraphics[width=0.7\textwidth]{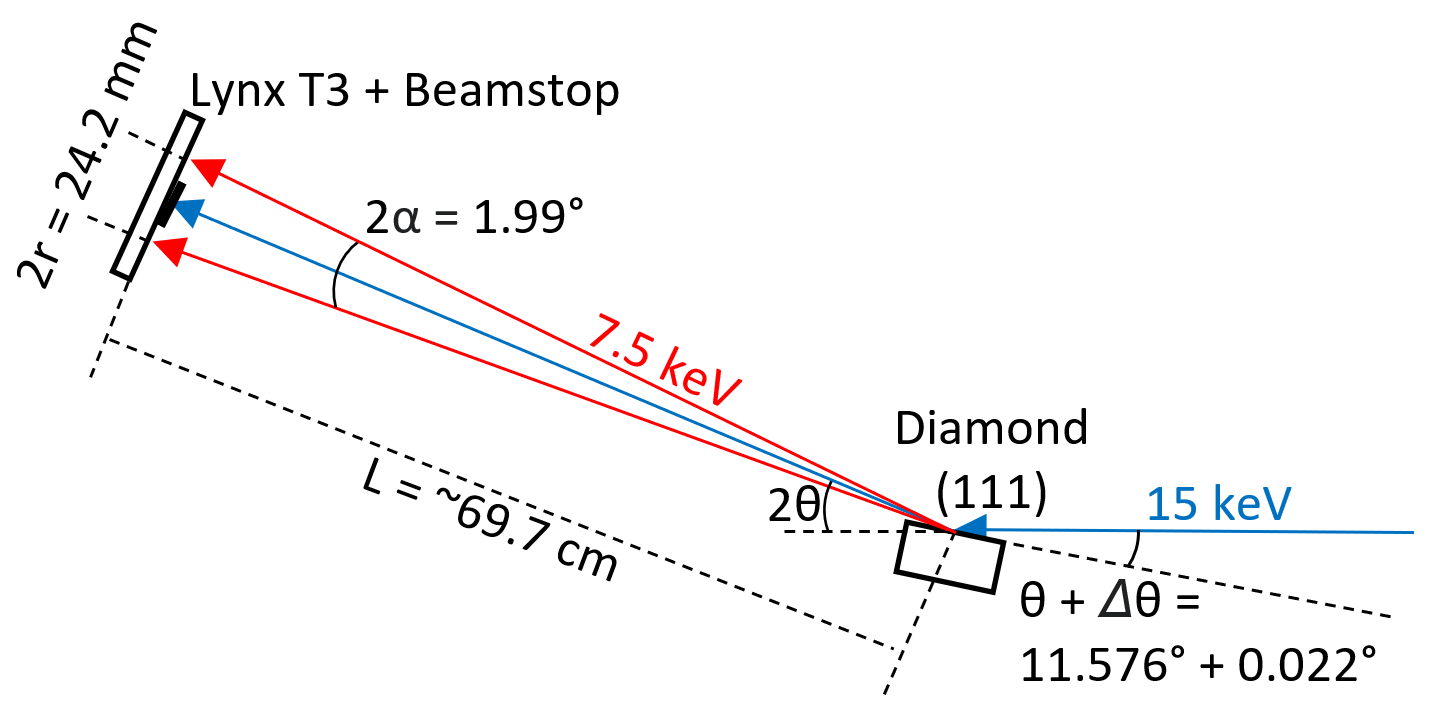}
\caption{\textbf{Schematic of the experimental setup employed at the CHX beamline.} A 15 keV partially coherent pump is incident upon the (111) diamond Bragg surface. A diffracted beam appears at the 2$\theta$ position, with half-energy SPDC photons emerging at an angle $\alpha$, determined by the detuning angle $\Delta \theta$. Non-degenerate X-rays are also generated but not depicted. The schematic is not drawn to scale.}
\label{efig1}
\end{figure}
\FloatBarrier

\begin{figure}[htp!]
\centering
\renewcommand{\figurename}{Extended Data Fig.}
\includegraphics[width=0.7\textwidth]{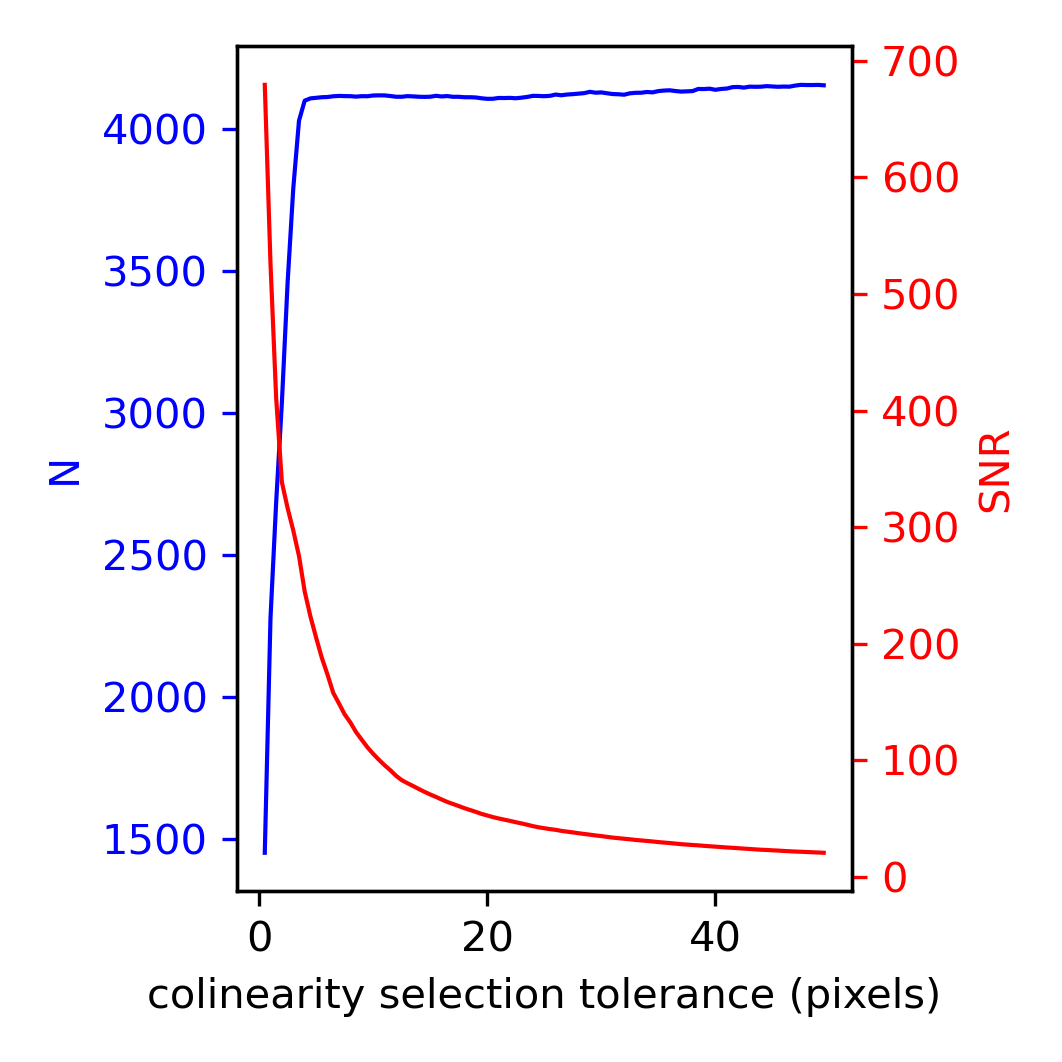}
\caption{
\textbf{Optimal collinearity selection parameter selection.} Collinearity of both single photons in the SPDC pair with the pump is required by conservation of momentum. The data are filtered to find colinear pairs with the goal of reducing background pairs. This filtering is performed by selecting only pairs within a small radius of the measured pump location. The SPDC count and SNR are plotted against the collinearity selection radius parameter. An optimal value of 4 pixels is found to maximize both the count and SNR.}
\label{efig2}
\end{figure}
\FloatBarrier

\begin{figure}[htp!]
\centering
\renewcommand{\figurename}{Extended Data Fig.}
\includegraphics[width=0.45\textwidth]{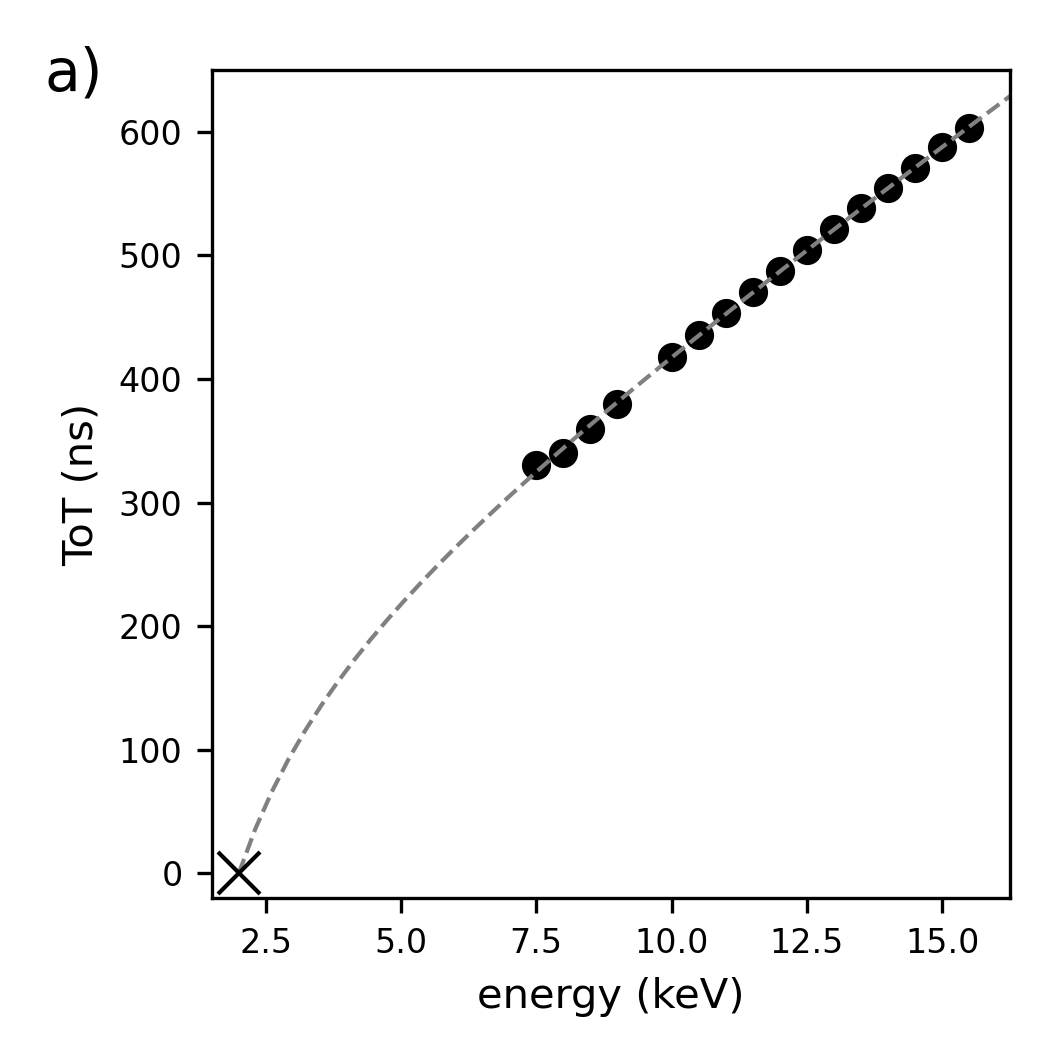}
\includegraphics[width=0.45\textwidth]{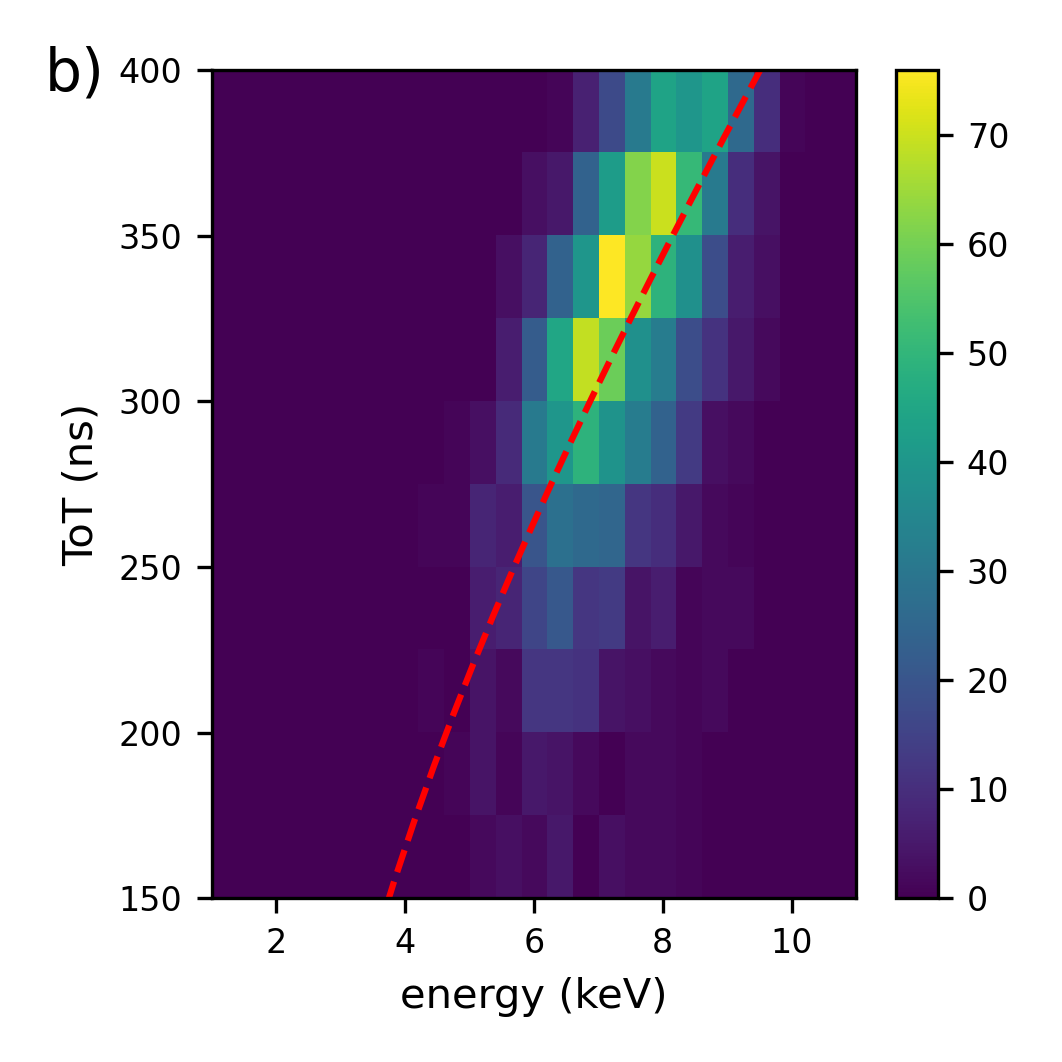}
\caption{
\textbf{Calibrations of X-ray energy from event ToT.} \textbf{a,} Calibrations from the average ToT as a function of energy from monochromatic X-ray scattering data (experimental data, black dots). The point (x marker) at (2 keV, 0 ns) was inferred from the technical specifications of the detector. A fit of the form $ToT(E) = aE + b - \frac{c}{E-t}$~\cite{Timewalk} was found using a = 40.0 ns/keV, b = 154 ns, c = 473 ns, and t = -0.19 keV, plotted as the dashed blue line. \textbf{b,} Histogram of measured SPDC single photon ToT vs. inferred energy from photon position using equation (3), with the overlayed calibration line from monochromatic X-ray calibrations.}
\label{efig3}
\end{figure}
\FloatBarrier

\begin{figure}[htp!]
\centering
\renewcommand{\figurename}{Extended Data Fig.}
\includegraphics[width=0.40\textwidth]{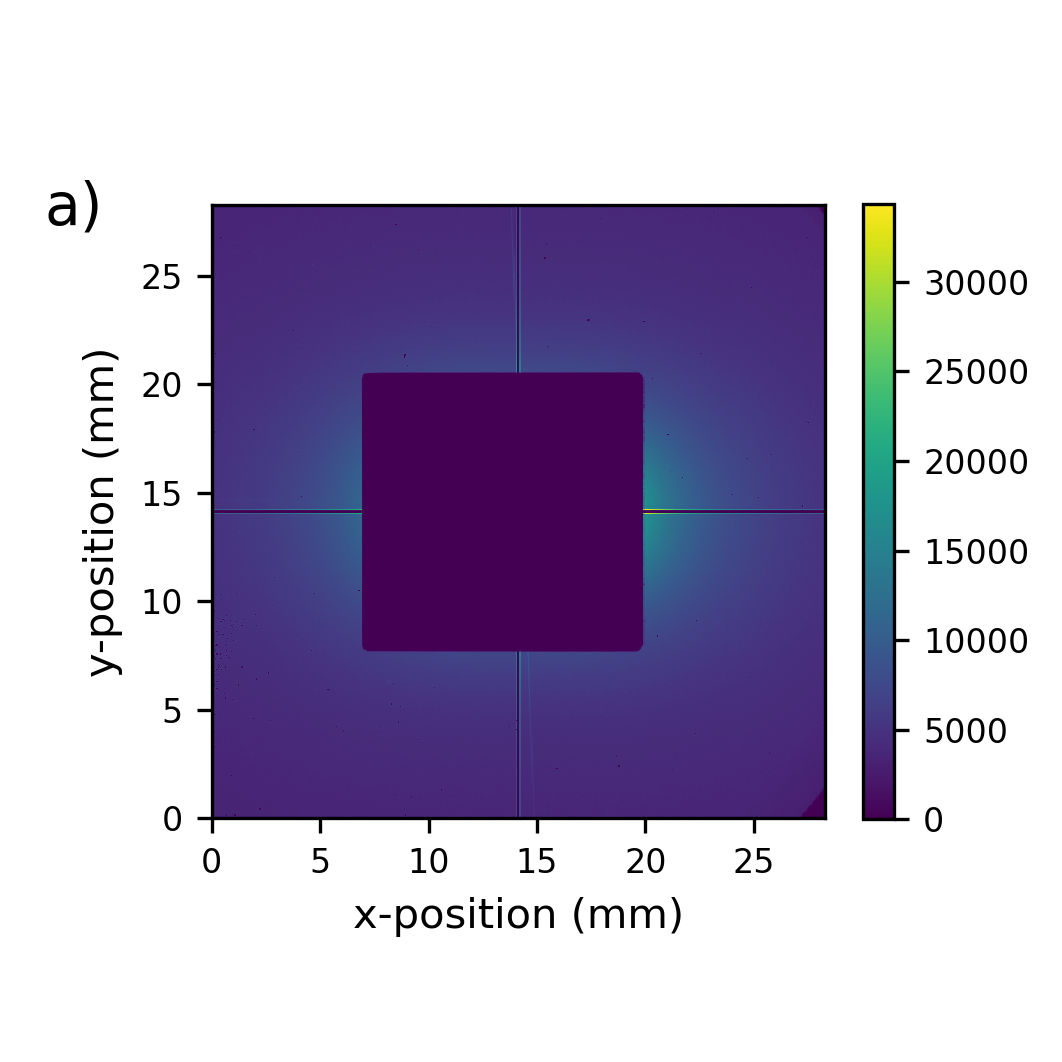}
\includegraphics[width=0.40\textwidth]{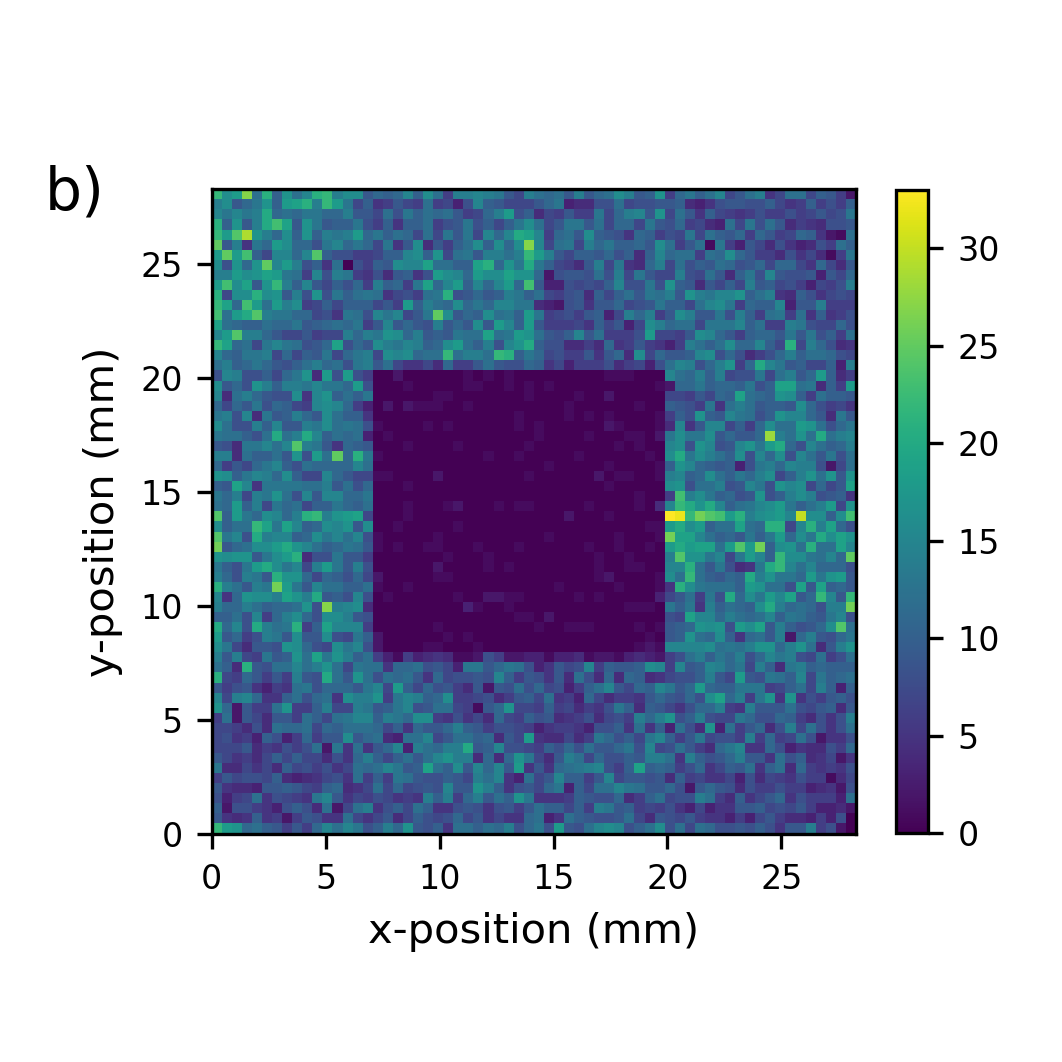}
\includegraphics[width=0.40\textwidth]{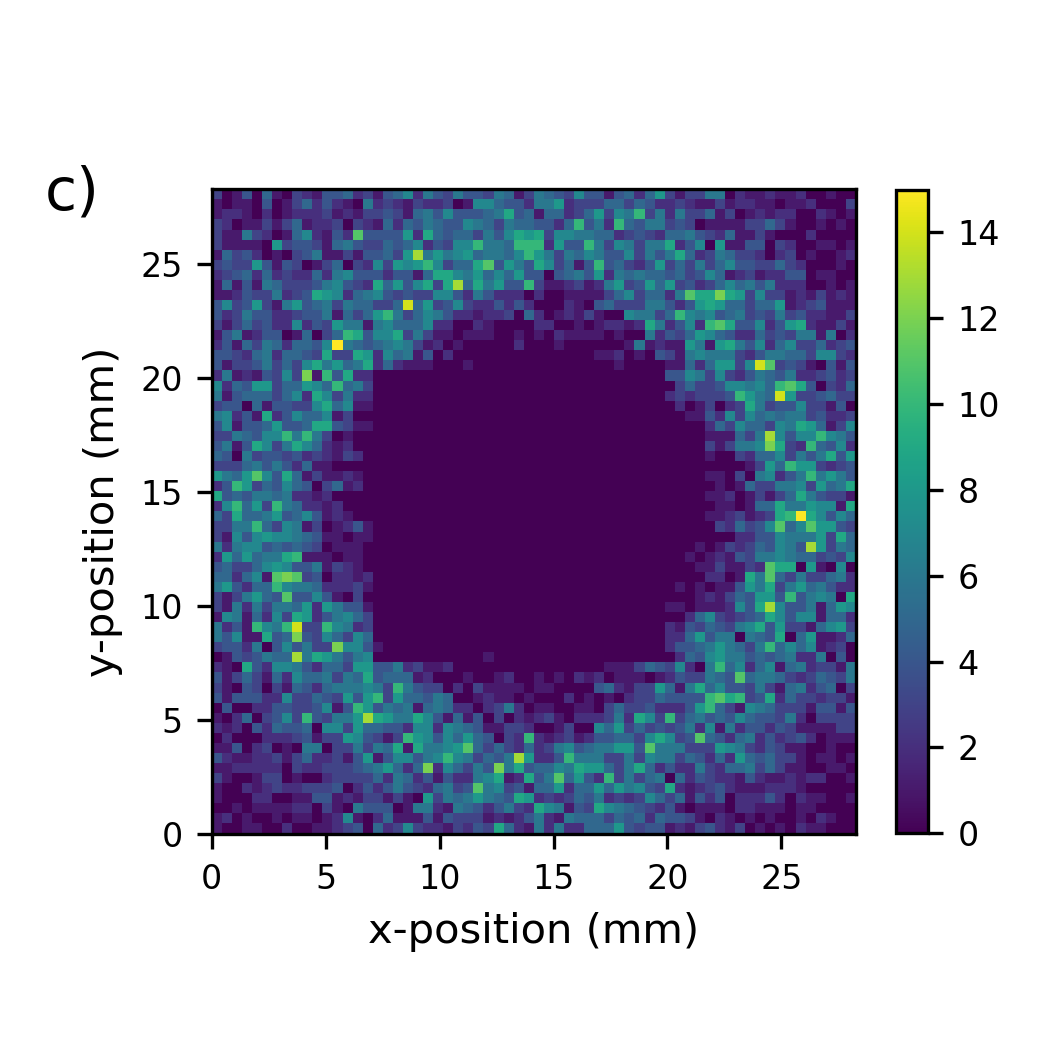}
\caption{\textbf{Measured X-ray data at various steps before the final selection.} \textbf{a,} Raw data from the detector from a one hour exposure prior to any filtering. \textbf{b,} SPDC photon pairs after ToT filtering and selecting events within a small coincidence window, without total energy or momentum selections. Significant levels of noise are indicated. \textbf{c,} SPDC photon pairs with total energy selections, but without momentum conservation. The noise level is reduced compared to (b) but can be further reduced with the momentum conservation selections.}
\label{efig4}
\end{figure}
\FloatBarrier

\section*{Supplementary Information}

\subsection*{Fitting Function}

The standardized normalized Gaussian (normal) function has the form:

\begin{equation}
g(x) = \frac{1}{\sigma \sqrt{2\pi}} \exp\left(-\frac{1}{2}\frac{(x-x_0)^2}{\sigma^2}\right)
\end{equation}

In our work, we fit the data of our $\Delta$t histograms (Fig. \ref{fig2}b) and $E_1 + E_2$ histogram (Fig. \ref{fig4}b) with a scaled and offset Gaussian curve, of the form:

\begin{equation}
g(x) = A \exp\left(-\frac{1}{2}\frac{(x-x_0)^2}{\sigma^2}\right) + d
\end{equation}

where \textit{d} is a constant offset. We calculate the number of pairs \textit{N} as:

\begin{equation}
N = \frac{A \sigma \sqrt{2 \pi}}{b}
\end{equation}

where \textit{b} is the experimental data histogram bin size.

The signal-to-noise ratio SNR is calculated as:

\begin{equation}
SNR = \frac{A}{d}
\end{equation}

The function parameters (i.e. A, $x_0$, $\sigma$, d) were fit to the data using nonlinear least squares.  

\subsection*{Derivation of Photon Energy vs. Emission Angle}

Equation (3), which gives the relationship between an SPDC photon's energy $E_x$ and its emission angle R(x) (which we simply call E and $\alpha$ in the manuscript), is derived from equation (7) in Eisenberger and McCall's treatment~\cite{Eisenberger}:

\begin{equation}
R(x) = \sqrt{2 \Delta \theta \left(\frac{y}{x}\right) sin(2 \theta)}
\end{equation}

where x and y are the fractions of the pump energy for the two down-converted photons, i.e. $E_x = x \hbar \omega_{pump}$ and $E_y = y \hbar \omega_{pump}$, or alternatively $x = \frac{E_x}{E_{pump}}$ and $y = \frac{E_y}{E_{pump}}$. Noting that energy conservation requires $y = 1 - x$, and thus $\frac{y}{x} = \frac{1-x}{x} = \frac{E_{pump}}{E_x} - 1$, by substitution and solving for $E_x$ one arrives at:

\begin{equation}
E_{x} = \frac{E_{pump} }{\frac{R(x)^2}{2 \Delta\theta \sin 2\theta} + 1}
\end{equation}

Dropping the subscript x and replacing R(x) with $\alpha$, for a relationship that applies for either photon in the pair, we arrive at our equation (3):

\begin{equation}
E = \frac{E_{pump}}{\frac{\alpha^2}{2 \Delta\theta \sin 2\theta} + 1}
\end{equation}

which may also be written as a function of the position \textit{r} from the 2$\theta$ position as:

\begin{equation}
E(r) =  \frac{E_{pump}}{\frac{\arctan^2(r/L)}{2 \Delta\theta \sin 2\theta} + 1}
\end{equation}

This final form was used for the conversion of the spatial location of SPDC single photons into energies (Fig. \ref{fig4}, Fig. \ref{fig6}, Extended Data Fig. \ref{efig3}).

\subsection*{Derivation of SPDC Efficiency}

The work of Eisenberger and McCall writes the cross-section of the process in their equation (8)~\cite{Eisenberger}:

\begin{equation}
\frac{d\sigma}{d\Omega} \propto (R(x) + R(y))^{-2}
\end{equation}

By substituting in the expressions for \textit{R(x)} and \textit{R(y)} as a function of a down-converted photon's energy fraction \textit{x}, one obtains:

\begin{equation}
\frac{d\sigma}{d\Omega} \propto \frac{1}{\left(\sqrt{a \frac{(1-x)}{x}}+\sqrt{a \frac{x}{1-x}}\right)^2}
\end{equation}

where \textit{a} is the factor $2\Delta\theta \sin 2\theta$ (not a function of x). Simplifying reveals:

\begin{equation}
\frac{d\sigma}{d\Omega} \propto \frac{x(1-x)}{2\Delta\theta \sin 2\theta}
\end{equation}

which indeed has the form of an inverted parabola with a maximum at \textit{x} = 0.5 and zeros at \textit{x} = 0 and 1.

\backmatter


\bmhead{Acknowledgments}

NSLS-II, a national user facility at Brookhaven National Laboratory (BNL), is supported in part by the U.S. Department of Energy, Office of Science, Office of Basic Energy Sciences Program under contract number DE-SC0012704. This project is supported through DOE-BER, Bioimaging Science Program KP1607020, within Biological Systems Science Division BSSD's Biomolecular Characterization and Imaging Science portfolio. It was also supported in part by the U.S. Department of Energy, Office of Workforce Development for Teachers and Scientists (WDTS) under the Science Undergraduate Laboratory Internships Program (SULI), and by the BNL Physics Department under the BNL Supplemental Undergraduate Research Program (SURP). This work was supported by the U.S. Department of Energy QuantISED award and BNL LDRD grants 19-30 and 22-22. 

We would like to recognize: Bryan Marino and Rick Greene, from NSLS-II's Complex Scattering Support team, for their assistance in fabricating our detector windows, beamstops, and their general support in maintaining beamline operations; John Lara, from NSLS-II's Structural Biology Support team, who made the crystal holders and the miniature goniometer support; Erik Muller, from the BNL Instrumentation Division, for loaning the diamond crystal used in our study; Erik Maddox, from Amsterdam Scientific Instruments, for providing useful information about and driver updates for the Lynx T3 detector; and F. C. Cruz for help with Fig. 1(a).

\section*{Declarations}

Authors declare no conflict of interest.




\bmhead{Author Contributions}

J.C.G., L.B., and A.F. performed the experiments. J.C.G., R.M., J.H., M.D., and A.N. performed data analysis. J.C.G. and M.D. performed supporting simulations. J.C.G., R.A.A., A.N., C.D., and S.M. prepared the manuscript. J.C.G., R.M., J.H., R.A.A., S.K., L.B., A.F., A.N., C.D., and S.M. discussed results. J.C.G., K.G., T.C., and D.A. contributed to the development of experimental control software. L.B., A.F., A.N., C.D., and S.M. conceived the experiment, provided experimental oversight, and secured the funding.

\bibliography{sn-bibliography}

\end{document}